\documentclass[fleqn,usenatbib]{mnras}

\usepackage{newtxtext,newtxmath}

\usepackage[T1]{fontenc}

\DeclareRobustCommand{\VAN}[3]{#2}
\let\VANthebibliography\thebibliography
\def\thebibliography{\DeclareRobustCommand{\VAN}[3]{##3}\VANthebibliography}

\usepackage{graphicx}	
\usepackage{amsmath}	

\title[DG Tau Streamer]{Cloudlet Capture Model for the Accretion Streamer onto the disk of DG Tau}

\author[T. Hanawa et al.]{
Tomoyuki Hanawa,$^{1}$\thanks{E-mail: hanawa@faculty.chiba-u.jp (TH)}
Antonio Garufi,$^{2}$
Linda Podio,$^{2}$ Claudio Codella,$^{2}$
and Dominique Segura-Cox$^{3}$
\\
$^{1}$Centre for Frontier Science, Chiba University,
1-33 Yayoi-cho, Inage-ku, Chiba 263-8522, Japan\\
$^{2}$INAF, Osservatorio Astrofisico di Arcetri, Largo E. Fermi 5, I-50125, Firenze, Italy\\
$^{3}$Department of Astronomy, The University of Texas at Austin, 2515 Speedway, Stop C1400, Austin, Texas 78712-1205, U.S.A.
}

\date{Accepted 2024 January 30. Received 2024 January 26; in original form 2023 December 2}

\pubyear{2023}

\begin{document}
\label{firstpage}
\pagerange{\pageref{firstpage}--\pageref{lastpage}}
\maketitle

\begin{abstract}
DG Tau is a nearby T Tauri star associated with a collimated jet, a circumstellar disk and a streamer a few hundred au long. The streamer connects to the disk at $\sim$50 au from DG
Tau. At this location SO emission is observed, likely due to the release of sulphur from dust grains caused by the shock of the impact of the accretion streamer onto the disk.
We investigate the possibility that the DG Tau streamer was produced via cloudlet capture on the basis of hydrodynamic simulations, considering a cloudlet initiating infall at 600 au from DG Tau with low angular momentum so that the centrifugal force is smaller than the gravitational force, even at 50 au. The elongation of the cloudlet into a streamer is caused by the tidal force when its initial velocity is much less than the free-fall velocity.  The elongated cloudlet reaches the disk and forms a high density
gas clump.  Our hydrodynamic model reproduces the morphology
and line-of-sight velocity of CS ($5-4$) emission from the Northern streamer observed with ALMA. We discuss the conditions for forming
a streamer based on the simulations.  We also 
show that the streamer should perturb the disk after impact for several thousands of years.
\end{abstract}



\section{Introduction} \label{sec:intro}

Young stellar objects are associated with disks of gas and dust.
Disks are formed since the early stages of the star formation process due to the conservation of angular momentum of the slowly rotating collapsing core \citep[e.g.][]{shu93}. To understand the evolution and the accretion and ejection phenomena onto/from the disk is of key importance as disks are the birthplace of planets.
This accretion disk theory assumes that the system is essentially symmetric around the rotation axis and that evolution is driven by angular momentum transfer.  Though this theory has succeeded in describing star formation in a general sense, it cannot explain asymmetric features discovered recently with high-resolution observations by the Atacama Large Millimeter/submillimeter Array \citep[ALMA; e.g.,][for L1489, TMC-1A, Per-emb-2, and GSS30-IRS5, respectively]{yen14,sakai16,artur19,pineda20}.

Gas accretion onto the disk is an important 
phenomenon as it may affect the mass, density, and chemical evolution of the disk. 
Asymmetric non-Keplerian gaseous structures, {\bf named accretion streamers}, have routinely been detected by ALMA and high-contrast imagers \citep[see, e.g., the review by][]{pineda23}. Kinetic modeling is required to confirm that asymmetric structures are indeed infalling streamers \citep[e.g.][]{thieme22,valdivia-mena22}. Some of the most notable examples of streamers connected to the disk are found around DO Tau \citep{huang22}, HL Tau \citep{yen19,garufi22}, SU Aur \citep{ginski21}, DG Tau \citep{garufi22}, and CB 68 \citep{kido23}.
Streamers may also feed material to evolved disks (less-embedded in their surrounding environments) in the form of late-stage infall events \citep{gupta23,huang23}, where the infalling material is more-often comprised of gas initially unbound to the initial dense core \citep{kuffmeier23}.
These findings are pivotal to understand how the disk accretion through streamers and the disk/environment interaction proceed throughout the disk evolution.

\begin{figure}
\includegraphics[width=\columnwidth]{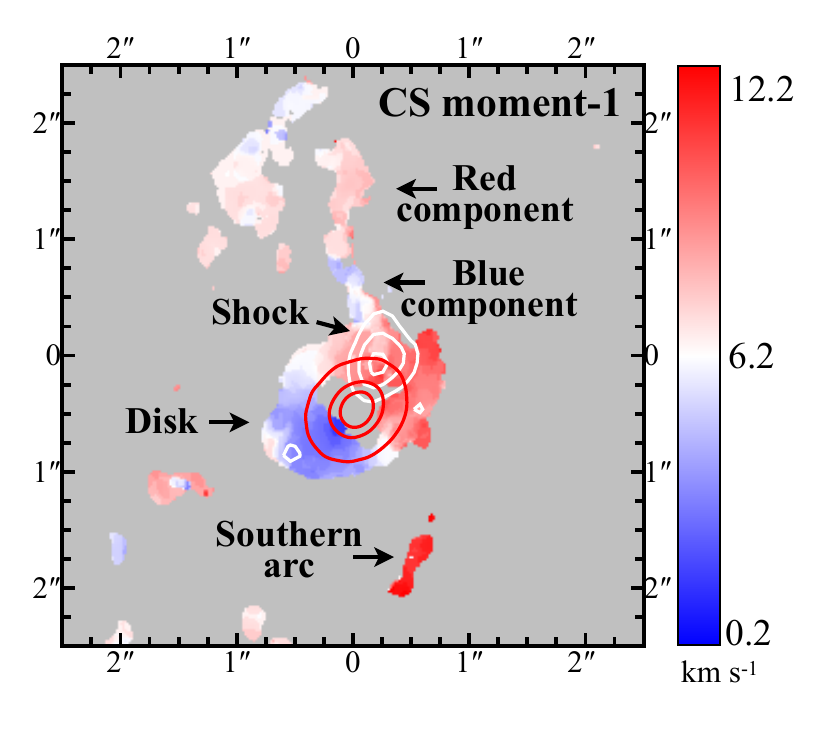}
\caption{The color shows the moment 1 of CS~$5-4$ emission line. 
The red and white contours denote the continuum at 1.3 mm and SO $4_5-3_4$ integrated intensity, respectively. Adopted from Fig. 2d) of \citet{garufi22}.\label{fig:map}}
\end{figure}

In this paper, we study a particular streamer known to accrete the disk of DG Tau, a T Tauri star located in the Taurus star forming region \citep[$d$=125 pc, Gaia data release 3,][]{gaia16,gaia23}. DG Tau is associated with a blueshifted jet \citep[e.g.,][]{eisloffel98}, a residual envelope observed at large scales \cite[e.g.,][]{kitamura96}, and a compact disk \citep[e.g.,][]{testi02,isella10}.
The streamer, recently observed in CS ($5-4$) and CO ($2-1$) emission in the context of the ALMA-DOT program \citep{garufi22}, extends for a few hundred au to the north of the disk as shown in Figure \ref{fig:map}. The northern streamer consists of red and blue components. The region where the streamer connects to the disk is associated with SO (white contours) and SO$_2$ emission, suggesting that an accretion shock occurs where the streamer impacts onto the disk, causing the release of sulphur from dust grain mantles due to dust grain sputtering and shattering \citep{neufeld89a,neufeld89b}. The enhancement of the emission from S-bearing molecules in shocks is commonly observed along jets and outflows \citep[e.g.,][]{bachiller97,lee10, codella14,podio21}, and at the interface between the infalling envelopes and disks \citep[accretion shocks, e.g.,][]{sakai14,oya16}.

The flat spectrum of DG Tau indicates the existence of hollow cavities outside the disk \citep[see, e.g.][]{whitney03}.  The morphology of the system suggests that the northern streamer resides in a narrow region in the northern cavity.  We surmise an invisible tenuous gas confines the streamer thin by pressure.  The tenuous gas component prevents denser gas from accreting and keeps the cavity transparent.  It should be warm since it has significant enough gas pressure to confine the streamer, despite the low density of the tenuous gas.  We assume that it is static for simplicity, since the streamer resides near the disk and far from the outflow axis.  The streamer is likely to come from a place distant from DG Tau and changes its shape when approaching DG Tau.  We assume the streamer was a dense gas clump in the surrounding molecular cloud.  It changed its trajectory several thousand years ago and began to approach DG Tau.   We name it \lq \lq cloudlet\rq\rq in this paper.  The cloudlet consists of cold molecular gas and inherits the chemical composition of the natal molecular cloud.   Our model considers the Keplerian disk rotating around DG Tau, the cloudlet, and the tenuous gas around the disk. Since we do not see any interaction of the streamer with the bipolar jets and infalling envelope, we do not include them in our model for simplicity.

When considering gas accretion from the envelope to the central star, we can employ analytic models \citep[e.g.,][]{ulrich76,cassen81} or the ballistic approximation \citep[see, e.g.,][]{sakai14}. Such models typically consider stationary inflows of material that are symmetric about the rotation axis, but can be leveraged to model asymmetric infall. Effects of infall that vary with time are not considered in these analytic models. These models assume that gas follows a free-fall orbit until it reaches the disk midplane. In other words, the models take into account 
hydrodynamic effects only after the accreting gas reaches the midplane. This approximation looks valid since the sound speed of the gas is much lower than the infall velocity near the star.  However, the gas pressure works to disperse an accreting gas cloudlet into a diffuse cloud if the cloudlet localized in a small region.  When discussing the formation of a streamer, we should examine the hydrodynamic effects since streamers should be confined against the gas pressure.

We aim to make a numerical simulation
reproducing the morphology and kinematics of the DG Tau streamer revealed in CS ($5-4$) emission. Though \cite{garufi22} reported two possible streamers, we concentrate on the northern one, which consists of red-shifted and blue-shifted component as shown in Figure \ref{fig:map}. Note that the red-shifted arc-like $^{12}$CO emission to the south discovered by \cite{gudel18} does not coincide with the CS ($5-4$) streamer on the sky, nor was it found to be kinematically-consistent with infall \citep{garufi22}.  The southern arc $^{12}$CO emission is likely to be a part of the outflow and not included in our modeling. Our model is similar to the cloudlet capture model for TMC-1A \citep{hanawa22}, although these two objects show different morphologies.
TMC-1A shows highly asymmetric features of a few 100 au scale 
in the CS ($5 - 4$) emission line. The blue-shifted component
is much stronger than the red-shifted one. 

In our modeling, we focus on the formation mechanism of an elongated structure.  Although the tidal force is the main driver, that alone is not enough.  \cite{kuffmeier20} considered capture of a cloudlet by a young star as a seed of second-generation disk. The cloudlet evolves not into a streamer falling onto the star, but an arc surrounding it.  Their cloudlet expanded appreciably during the infall since they considered only one component of gas. We assume that a warm neutral gas surrounding DG Tau confines an infalling cloudlet to form a streamer. We aim to follow the infall of cloudlet until its head reaches the disk.

The disk of DG Tau has been recently imaged with ALMA in  continuum emission and in several molecular tracers \citep{gudel18,podio19,podio20}. These studies constrained the disk inclination ($i=35^\circ$) and position angle (PA=135$^\circ$). The disk rotation (with the NW side redshifted and the SE blushifted) and the PA of the blueshifted jet \citep[PA=225$^\circ$,][]{eisloffel98} indicates that the disk rotates clockwise and the near-side is to the NE.
\cite{garufi22} obtained a best fit elliptic orbit for the streamer under the ballistic approximation, i.e., by taking into account only the gravity of DG Tau.  The orbit is 10$^\circ$ inclined with respect to the disk plane and above the disk plane in the  northwest. They assumed that the streamer originated from 450 au away from DG Tau with small radial (0.4~km~s$^{-1}$) and rotation (0.13~km~s$^{-1}$) velocities.  We consider a cloudlet taking a similar orbit and examine how it evolves.

This paper is organized as follows. We describe our model and methods in \S 2.  We show the results of 3 selected models in \S 3. 
We show the observed channel maps of CS ($5-4$) emission of the streamer and compare them with the output of the hydrodynamic model in \S 4. We discuss the condition for the formation of a streamer in \S 5.  
We summarize our results in \S 6.  

\section{Model}

This section describes our modeling and methods of computation. We introduce the basic equations and numerical methods for solving them in \S\S 2.1. We show the initial model of a cloudlet infalling onto a disk rotating around DG Tau in \S\S 2.2. We outline the methods of the mock observation in \S\S 2.3.

\subsection{Basic Equations}
We use the hydrodynamical equations,
\begin{eqnarray}
\frac{\partial \rho}{\partial t} + \mathbf{\nabla} \cdot
\left( \rho \mathbf{v} \right) & = & 0, \label{mcon} \\
\frac{\partial}{\partial t} \left( \rho \mathbf{v} \right)
+ \mathbf{\nabla} \cdot \left( \rho \mathbf{vv} + P \mathbf{I} \right) & = & 
- \rho \mathbf{\nabla} \Phi , \label{Pcon}  \\
\frac{\partial}{\partial t}( \rho E ) + \mathbf{\nabla} \cdot
\left[ \left( \rho E + P \right) \mathbf{v} \right] & = & 
- \rho \mathbf{v} \cdot \mathbf{\nabla} \Phi , 
\label{energy} \\
E = \frac{|\mathbf{v}| ^2}{2} + \frac{P}{\rho (\gamma - 1)} , & & 
\end{eqnarray}
to describe gas accretion onto a protostar associated with a gas disk.
The symbols, $ \rho $, $ P $, and $\mbox{\boldmath$I$}$ denote the density, pressure, the unit tensor, respectively, 
while $\mathbf{v} $ and $ \Phi $ denote the velocity and gravitational potential, 
respectively.  The gas pressure is expressed as
\begin{eqnarray}
P & = & \frac{k}{\bar{m}} \rho T , 
\end{eqnarray}
where $ T $, $ k $, and $ \bar{m} $ denote the temperature, 
Boltzmann constant, and mean molecular weight, respectively.
Equation (\ref{energy}) considers neither heating nor cooling explicitly.
However, the specific heat ratio is taken to be $ \gamma = 1.05 $ so that
the temperature remains nearly constant.  This approximation mimics the
situation where both the molecular and atomic gases remain nearly isothermal.

We consider three gas components: cloudlet, disk, and warm neutral medium.
The cloudlet and disk consist of molecular gas while the warm neutral medium
consists of of atomic gas.  The mean molecular weight is assumed to be
$ \bar{m} _{\rm c} = \bar{m} _{\rm d} = 2.3~m_{\rm H} $ for the cloudlet and disk, and
$ \bar{m} _{\rm w} = 1.17~m_{\rm H} $ for the warm neutral medium, where
$ m _{\rm H} $ denotes the mass of a hydrogen atom.

We introduce a variable, 
\begin{equation}
c = \left\{ 
\begin{array}{ll}
1 & \mbox{(cloudlet)} \\
0 & \mbox{(warm neutral gas)} \\
-1 & \mbox{(disk)} \\
\end{array}
\right. ,
\end{equation}
to trace the three components.  For this purpose, we
solve 
\begin{eqnarray}
\frac{\partial}{\partial t} \left( c \rho \right)
+ \mathbf{\nabla} \cdot \left( c \rho \mathbf{v} \right)  =  0 ,
\label{ctrace}
\end{eqnarray}
simultaneously.  Equation (\ref{ctrace}) means that $ c $ remains
constant along the gas element, i.e., the Lagrangian derivative,
$ Dc/dt = 0 $, vanishes.

Using the position vector $ \mathbf{r} $, we approximate the gravity to be
\begin{eqnarray}
- \mathbf{\nabla} \Phi & = & 
- \frac{GM \mathbf{r}}{(\max{ |\mathbf{r}|, a}) ^3}, \label{gravity}
\end{eqnarray}
where $ G $, $ M $ and $ a $ denote the gravitational constant, the mass
of the central star, and the unit length, respectively.  Equation (\ref{gravity})
means that we consider only the central star, with the mass taken
to be $ M = 0.7~\mbox{M}_\odot$, as a source of the gravity. The gravity is
reduced artificially in the region of $ | \mathbf{r} | \le a = 30~\mbox{au} $
so that we can avoid numerical difficulties.
The corresponding gravitational potential is expressed as
\begin{eqnarray}
\Phi & = & \left\{ \begin{array}{ll} - \displaystyle \frac{GM}{|\mathbf{r}|}
& (|\mathbf{r}| > a)  \\
- \displaystyle \frac{GM}{2 a ^3} \left( 3 a ^2 - | \mathbf{r} | ^2 \right) & (|\mathbf{r}| \le a) \\
\end{array}
\right. 
\end{eqnarray}
We take into account neither magnetic fields nor turbulent viscosity, for simplicity.

The mass of DG Tau was assumed to be $0.3~\mbox{M} _\odot $ in the modeling
of streamer in \cite{garufi22}.  However, the mass of DG Tau is still uncertain. \cite{testi02} derived $ M = 0.67 ~\mbox{M} _\odot$ from $^{13}$CO emission from the disk by assuming Keplerian rotation and an inclination of 38$^\circ$. \cite{podio13} confirmed that water vapor emission line profile is consistent with $0.7~\mbox{M} _\odot$. Hence, we adopt 0.7~M$_\odot$ in this paper to
model both disk and streamer.  Since the CS emission from the streamer is weak,
we can constrain only weakly the mass from its line-of-sight velocity.

We use cylindrical coordinates, $ (r, \varphi, z) $, in our
numerical simulations.  The central star of DG Tau resides at the
origin while the disk is in the plane of $ z = 0 $.  We cover the
cylindrical volume of $ r \le R _{\rm out} $ and $ |z| \le Z _{\rm out} $.
The spatial resolution is constant at $ r \Delta \varphi \simeq dr = dz =\Delta r _0 $
around the origin.
Outside the central cylindrical region, the spatial resolution is $ \Delta r / r = \Delta z /|z| = \Delta \varphi = 360 ^\circ/ N _\varphi $,
where $ N _\varphi = 384 $ denotes the maximum number of numerical cells in the
azimuthal direction. The parameters are set
to be $ \Delta r _0 = 1.5~\mbox{au} $ in model B and 0.75 au  in the rest of the models.  See \cite{hanawa21} for further
details of the numerical code.

\subsection{Initial Model}

We assume the warm gas in the cavity surrounding the disk is isothermal at $ T _{\rm w} $ and in the hydro-static equilibrium for simplicity, although we note that realistically the environment surrounding DG Tau has a temperature gradient and non-zero initial motions. Then, the pressure is expressed as
\begin{eqnarray}
P _{\rm w} (\mathbf{r}) & = & P _0 \exp 
\left[ - \frac{\bar{m} _{\rm w} \Phi (\mathbf{r})}{k T _{\rm w}} \right] , \label{warm-init}
\end{eqnarray}
where $ P _0 $ denotes the pressure at a large distance from DG Tau.
We set the temperature to be $ GM\bar{m} _{\rm m}/ a k T _{\rm w} = 0.45 $ so that the warm gas is weakly bound to DG Tau at the distance of 30 au.
It is $ T  _{\rm w} = 1.42 \times 10 ^3~\mbox{K} $ in the physical unit for $ \bar{m} _{\rm w} = 1.17~m_{\rm H} $ and $ M = 0.7~M_\odot $. Our estimate of $ T _{\rm w} $ is based on the theoretical consideration of the thermal equilibrium of the interstellar medium \citep[see, e.g.,][]{inoue06}. When the main heating source is cosmic rays and the main coolant is ionized carbon, the temperature of the warm component is $ 1000-2000 $~K at a high pressure while that of the cold component is below 100 K.
Accordingly the number density is expressed as
\begin{eqnarray}
n _{\rm w} (\mathbf{r}) & = & \frac{P _0}{k T _{\rm w}} 
\exp \left[ - \frac{\bar{m} _{\rm w} \Phi (\mathbf{r})}{k T _{\rm w}} \right] , 
\end{eqnarray}
and proportional to $ P _0 $.  Since Equations (\ref{mcon}) through (\ref{energy}) are proportional to the density, our model has no specific density.  We will define the parameter $ P _0 $ later, but the choice of $ P _0 $ does not alter the velocity. 

The cloudlet, {\bf the precursor of the streamer}, is a gas sphere having the radius $ R _{\rm c} $
and centered at $ \mathbf{r} _0 $ at the initial stage.
We specify the initial cloudlet center by
\begin{eqnarray}
\mathbf{r} _0 & = & r _0 \cos \theta _0 \cos \varphi _0 \mathbf{e} _x 
+ r _0 \sin \varphi _0 \mathbf{e} _y \nonumber \\
& & - r _0 \sin \theta _0 \cos \varphi _0 
\mathbf{e} _z , 
\end{eqnarray}
where $ \mathbf{e} _x $, $ \mathbf{e} _y $, and $ \mathbf{e} _z $ denote 
the $ x $- , $ y $- and $ z $-components
of the unit vectors in Cartesian coordinates. 

We consider two cases of orbits, elliptic and parabolic.
In both the cases, the initial velocity distribution is
expressed as
\begin{eqnarray}
\mathbf{v} _0 & = & v _{0,r} \frac{\mathbf{r} _0}{\left|\mathbf{r} _0\right|} 
+ v _{\varphi, 0} \mathbf{e} _z ^\prime \times
\frac{\mathbf{r} _0}{\left|\mathbf{r} _0\right|} , \\
\mathbf{e} _z ^\prime & = & \sin \theta _0 \mathbf{e} _x + 
\cos \theta _0 \mathbf{e} _z 
\end{eqnarray}
such that the parameter $ \theta _0 $ denotes the inclination of the orbital plane 
to the $  z $-axis.
We specify the rotation velocity by
\begin{eqnarray}
v _{\varphi,0} & = & - \frac{\sqrt{G M r _{\rm cen}}}
{\left| \mathbf{r} _0 \right|} , \label{c_rotation}
\end{eqnarray}
where $ r _{\rm cen}$ denotes the centrifugal radius, i.e.,
the radius at which the centrifugal force balances with the
gravitational force. We arrange $ \varphi _0 $ so that the periastron of the orbit is located in the plane of $ x = 0 $ and $ y < 0 $.

\begin{figure}
\includegraphics[width=\columnwidth]{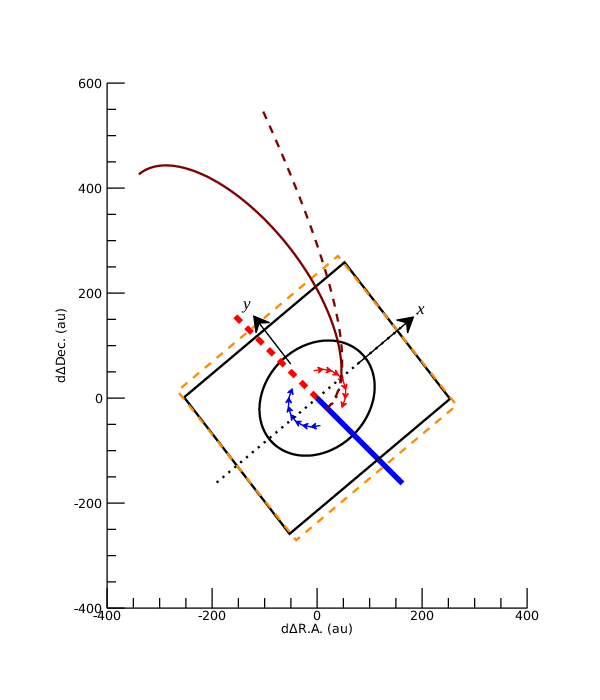}
    \caption{This diagram denotes the geometry of the elliptic and parabolic orbits of the cloudlet by the solid and dashed curves, respectively. The black ellipse and square denote the disk and disk plane, respectively. The blue solid and red dashed lines denote the blue-shifted and red-shifted jets, respectively. The disk rotates clockwise on the sky plane as shown by the blue and red arrows. The black arrows denote the $ x $-and $ y $-axis in  Cartesian coordinates in the numerical computation.  The orbit of the cloudlet, shown by the orange arc, is inclined by 10$^\circ$ from the disk plane.}
    \label{fig:orbit}
\end{figure}

When considering an elliptic orbit, we set the cloudlet at
the apoastron ($\varphi = 180^\circ$) at the initial stage.  
Accordingly, the
initial radial velocity vanishes ($ v _{r,0} = 0 $).
The eccentricity of the orbit is evaluated to be
\begin{eqnarray}
\varepsilon & = & 1 - \frac{r _{\rm cen}}{r _0} .
\end{eqnarray}
The orbit intersects with the disk midplane ($ z = 0 $) at 
$ (r, \varphi) = (r _{\rm cen},~0^\circ) $ and
has the periastron distance,
\begin{eqnarray}
r _{\rm min} & = & \frac{ r _0 (1 - \varepsilon)}{1 + \varepsilon} .
\end{eqnarray}
Then, the semi-major axis of the orbit and the orbital period are evaluated to be 
\begin{eqnarray}
a _{\rm maj} & = & \frac{r _0}{1 + \varepsilon} = 
\frac{r _0 {}^2}{2 r _0 - r _{\rm cen}} , \\
P _{\rm orb} & = & 6.12 \times 10 ^3 
\left( \frac{r _0}{600~\mbox{au}} \right) ^{3/2} 
\left( \frac{1 + \varepsilon}{2} \right)^{-3/2}~
\mbox{yr} ,
\end{eqnarray}
respectively.  
Figure \ref{fig:orbit} shows the geometry of the cloudlet and disk, the latter of which lies with an inclination of 35$^\circ $ and the near side in the northeast. The orbital plane of cloudlet is closer to us in the northeast than the disk plane. Thus, the cloudlet recedes from us when approaching DG Tau. We assume that the cloudlet is captured by turbulence at a distance of $> 500~\mbox{au}$ and turns into a highly elliptic or parabolic orbit.

When the cloudlet has a parabolic orbit, the initial radial velocity  is
expressed as 
\begin{eqnarray}
v _{r,0} (\mathbf{r}) & = & \left\{
- \sqrt{\displaystyle \frac{2GM}{\left|\mathbf{r} \right|}
- v _{\varphi,0} ^2}  
\right. .
\end{eqnarray}
The head of the cloudlet
has a higher infall velocity than the tail. In other words, the cloudlet head moves faster than the tail. Since the initial radial distance is 10 times larger than the centrifugal radius ($ r _0 = 10 r _{\rm cen} $), the eccentric anomaly is 154$^\circ$ at the initial stage in model C. The periastron is located at $ x $ = 0 and close to those of the elliptic orbits.  Thus the orbits are close each other near DG Tau.

We assume that the cloudlet has the same pressure as the surrounding warm neutral medium
so that it does not expand or shrink immediately.  The number density
inside the cloudlet is given by
\begin{eqnarray}
n _{\rm c} (\mathbf{r}) & = & \frac{P _0}{k T _{\rm c}}
\exp \left[ - \frac{\bar{m} _{\rm w} \Phi (\mathbf{r})}{k T _{\rm w}} \right] , 
\end{eqnarray}
for $ | \mathbf{r} - \mathbf{r} _0 | \le R _{\rm c} $, where
$ T _{\rm c} $ and $ R _{\rm c} $ denote the initial temperature and radius of the cloudlet, 
respectively.  
In the following, we assume $ G M \bar{m} _{\rm c} / r _0 k T _{\rm c} = 5.0 \times 10 ^{-3} $ so that the thermal energy is comparable with the gravitational energy at the distance of  6000 au ($ = 10~r _0$) from DG Tau. The temperature corresponds to $ T _{\rm c} = 28.6~\mbox{K} $ for $ \bar{m} _{\rm c} = 2.3~m _{\rm H} $ and $ M = 0.7~M_\odot $.  We assume assume $ P _0 / (k T _{\rm c}) = 10^4~\mbox{cm} ^{-3} $ though the density is scale free as mentioned earlier.  

\begin{figure}
\centering
\includegraphics[width=0.45\textwidth]{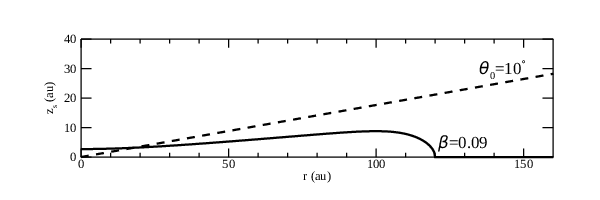}
    \caption{The solid curve denotes the half thickness of
    the disk as a function of the radial distance from the rotation axis for $ R _{\rm d} = 120~\mbox{au}$
    and $ \beta = 0.09 $.  
    The dashed line denotes the plane inclined 80$^\circ$ with respect to the
    disk axis ($r = 0$).}
    \label{disk-shape}
\end{figure}

We consider an isothermal disk rotating around the $ z $-axis for simplicity, though the disk should have radial and vertical temperature gradients. Considering the gradients is beyond the scope of this paper because our spatial resolution is limited. 
The velocity is expressed as
\begin{eqnarray}
\mathbf{v} & = & v _\varphi (r) \mathbf{e} _\varphi .
\end{eqnarray}
We assume that the disk is bounded in the region of $ | z | \le z _s (r)$,
\begin{eqnarray}
z _{\rm s} (r) & = & \beta \sqrt{(a ^2 + r ^2) 
\tanh \left[ \frac{2 \left( R _{\rm d} - r \right)}{a} \right] }  ,
\end{eqnarray}
where $ R _{\rm d} $ denotes the disk radius and
$ \beta $ specifies the disk thickness. 
Figure~\ref{disk-shape} shows  the hight of the disk upper boundary ($ z _s $) as
a function of $ r $ for $ R _{\rm disk} = 120~\mbox{au} $.
The pressure distribution is given by
\begin{eqnarray}
P _{\rm d} (r, z) & = & P _{\rm w} [r, z _{\rm s}(r)] \nonumber \\
& & \times \exp\left\{ \frac{\bar{m} _{\rm d}}{k T _{\rm d}} 
[\Phi (r, z_s) - \Phi (r, z)] \right\} , \label{pdisk}
\end{eqnarray}
where $ T _{\rm d} $ denotes the temperature of
the disk gas.  We assume $ G M \bar{m} _{\rm d} / (a k T _{\rm d}) = 0.01 $, i.e., $ T _{\rm d} = 57.3~\mbox{K} $.
Equation (\ref{pdisk})
describes hydrostatic equilibrium in the vertical ($z$) direction.
The rotation velocity is set to be
\begin{eqnarray}
v _\varphi (r)  & = & - \sqrt{\left( \displaystyle 1 - \frac{\bar{m} _{\rm w} T _{\rm d}}{\bar{m} _{\rm d} T _{\rm w}} 
\right) 
\left( \frac{\partial \Phi }{\partial r}
+ \frac{\partial \Phi}{\partial z} 
\frac{\partial z _{\rm s}}{\partial r}\right) } , \label{rotation}
\end{eqnarray}
where we evaluate $ \partial \Phi / \partial r $ and $ \partial \Phi /\partial z $
at the disk surface, $ z = z _s (r) $, so that the sum of the centrifugal and pressure forces
balance gravity. 

At the initial stage, we do not include the gas disk in early stages of simulation
($ 0 \le t \le t _{\rm d} $).
We refresh the density, velocity, pressure and the color field in the
region of $ r < R _{\rm d} ^\prime $ 
and $ | z | \le z _{\rm d} ^\prime $ 
to insert the gas disk at $ t = t _{\rm d} $.

Table~\ref{parameters} summarizes the parameters for the three models shown in this paper. We set the radius of the cloudlet, $ R _{\rm c}$ to reproduce the width and length of the streamer.

\begin{table*}
\caption{Model Parameters (see \S 2) \label{parameters} }
\begin{tabular}{cccccccccccc}
\hline
Model & $r _0$ & $ r _{\rm cen}$& $e$ & $|\mathbf{v} _0 |$
& $R _{\rm d} ^\prime $ &
$ Z _{\rm d} ^\prime $ & $t _{\rm d}$ & $R _{\rm out} $ &
$ Z _{\rm out} $ & $ \Delta r _0 $  \\
name & (au) & (au) & & (km~s$^{-1}$)  & (au) & (au) 
& (yr) & (au) & (au) &  (au) \\
\hline
A & ~600 & 60 & 0.90 & 0.32 & 181 & 18.4 & 2337 & ~871 & 213 & 0.75 \\
B & 1200 & 60 & 0.95 & 0.16 & 295 & 52.2 & 8181  & 1399 & 338 & 1.50 \\
C & ~600 & 60 & 1.00 & 1.44 & 181 & 18.4 & ~904 & ~871 & 213 &  0.75  \\
\hline
\end{tabular} \\
The temperature is set to be $1.42 \times 10 ^3 $~K for the warm neutrral gas, 28.6~K for the cloudlet and 57.3~K for the disk in all the models.
 \end{table*}



\subsection{Mock Observation}
In this section we briefly illustrate the methodology adopted to produce synthetic channel maps for comparison with observations.
We use the Cartesian coordinates,
\begin{eqnarray}
\mathbf{r} & = & X \mathbf{n} _1 + Y \mathbf{n} _2 + s \mathbf{n} _3 ,
\end{eqnarray}
to evaluate the emission expected from our model simulation.
Our line of sight is parallel to the unit vector, $ \mathbf{n} _3 $,
while $ \mathbf{n} _1 $ and $ \mathbf{n} _2 $ denote the unit vectors
in the directions of decreasing right ascension and increasing declination,
respectively. The line of sight is related to the cylindrical coordinates by
\begin{eqnarray}
\mathbf{n} _3 & = & \left( \begin{array}{c} - \sin i \sin \varphi _{\rm obs} \\
\sin i \cos \varphi _{\rm obs} \\ \cos i \end{array} \right) ,
\end{eqnarray}
where $ i $ and $ \varphi _{\rm obs} $ denote the disk inclination
angle and the observer's azimuth in the cylindrical coordinates. 
The other unit vectors are
expressed as
\begin{eqnarray}
\mathbf{n} _1 & = & \left( 
\begin{array}{c} \cos \chi _{\rm obs} \cos \varphi _{\rm obs} 
+ \cos i \sin \chi _{\rm obs} \sin \varphi _{\rm obs} \\
\cos \chi _{\rm obs} \sin \varphi _{\rm obs} 
- \cos i \sin \chi  _{\rm obs} \cos \varphi _{\rm obs} \\
\sin i \sin \chi _{\rm obs}  \end{array} \right) , \nonumber \\
\\
\mathbf{n} _2 & = & \left( \begin{array}{c} 
\sin \chi _{\rm obs} \cos \varphi _{\rm obs} -
\cos i \cos\chi _{\rm obs} \sin \varphi _{\rm obs} \\
\sin \chi _{\rm obs} \sin \varphi _{\rm obs}
+ \cos i \cos\chi _{\rm obs} \cos \varphi _{\rm obs} \\
- \sin i \cos \chi _{\rm obs}  \end{array} \right), \nonumber \\
\end{eqnarray}
where $ \chi _{\rm obs} $ denotes the orientation of the major axis of the disk  on the sky.
We measure the major axis counter-clockwise from $ \mathbf{n} _1 $ (West).
In the following we assume $ i = 35 ^\circ $ and $ \chi = 45 ^\circ $ to
reproduce the Doppler shift due to the disk rotation.

For simplicity, we assume that the intensity at a given radial
velocity is proportional to the gas density integrated over the
line of sight,
\begin{eqnarray}
\Sigma (X, Y; v) & = &
\int \frac{\rho (\mathbf{r})}{\sqrt{2\pi} \sigma}
\exp \left\{ - \frac{1}{2} \left[ \frac{v _{\rm los} - v}{\sigma} \right] ^2 \right\} ds ,
\nonumber \\
& & \\
\mathbf{r} & = & X \mathbf{n} _1 + Y \mathbf{n} _2 + s \mathbf{n} _3 , \\
v _{\rm los} & = & - \mathbf{v} \cdot \mathbf{n} _3 ,
\end{eqnarray}
where $ \sigma $ denotes the local line width in units of velocity. 
We assume $ \sigma = 0.1~\mbox{km~s} ^{-1} $ to obtain a smooth 
channel map.  For comparison with observation, we sum up the column density along
the line of sight,
\begin{eqnarray}
\bar{\Sigma} (X, Y; v) & = & \frac{1}{N}
\sum _{i=1} ^N \Sigma \left( X, Y; v _i \right) , \\
N & = & \frac{\Delta v}{\sigma} , \\
v _i & = & v + \left( i - \frac{N + 1}{2} \right) \sigma , 
\end{eqnarray}
over the velocity range of $ v - \Delta v/2 \sim v + \Delta v / 2 $.  
Since the velocity resolution is $ 0.6~\mbox{km~s} ^{-1} $ for
the CS ($ J =5-4$) emission from DG Tau, we show the average of
6 column density differentials along the line of sight for the
direct comparison with observations. We also show mock channel maps
with higher velocity resolution for our analysis.

We note that our mock observation does not take account of self-absorption, non-uniform excitation temperature, and chemical in-homogeneity. These factors should affect the expected intensity. However, it is beyond the scope of this work for us to include these factors since they are highly uncertain. The mock channel maps enable us to compare the model and observation kinematics directly since they provide morphology and line of sight velocity simultaneously.

\section{Results from the modeling}

\subsection{Model A}

First, we introduce model A as a reference.  The cloudlet has a radius of 90 au and is located at 600 au away from the star at the initial stage.  
The density of the cloudlet is nearly uniform and
in the range of $ 9.96 \times 10 ^5~\mbox{cm} ^{-3} < n < 1.02 \times
10 ^{6}~\mbox{cm} ^{-3} $ and  the mass density is 90 times higher than
that of the surrounding warm gas.  The density is slightly higher on the side of the cloudlet closer to the star. The mass of the cloudlet is $ 1.97 \times 10 ^{-5}~\mbox{M}_\odot $. The initial velocity is $0.21~\mbox{km~s}^{-1}$, and accordingly the motion is subsonic for the assumed temperature of $ T _{\rm c} = 28.6~\mbox{K} $. 

\begin{figure*}
\includegraphics[width=\columnwidth]{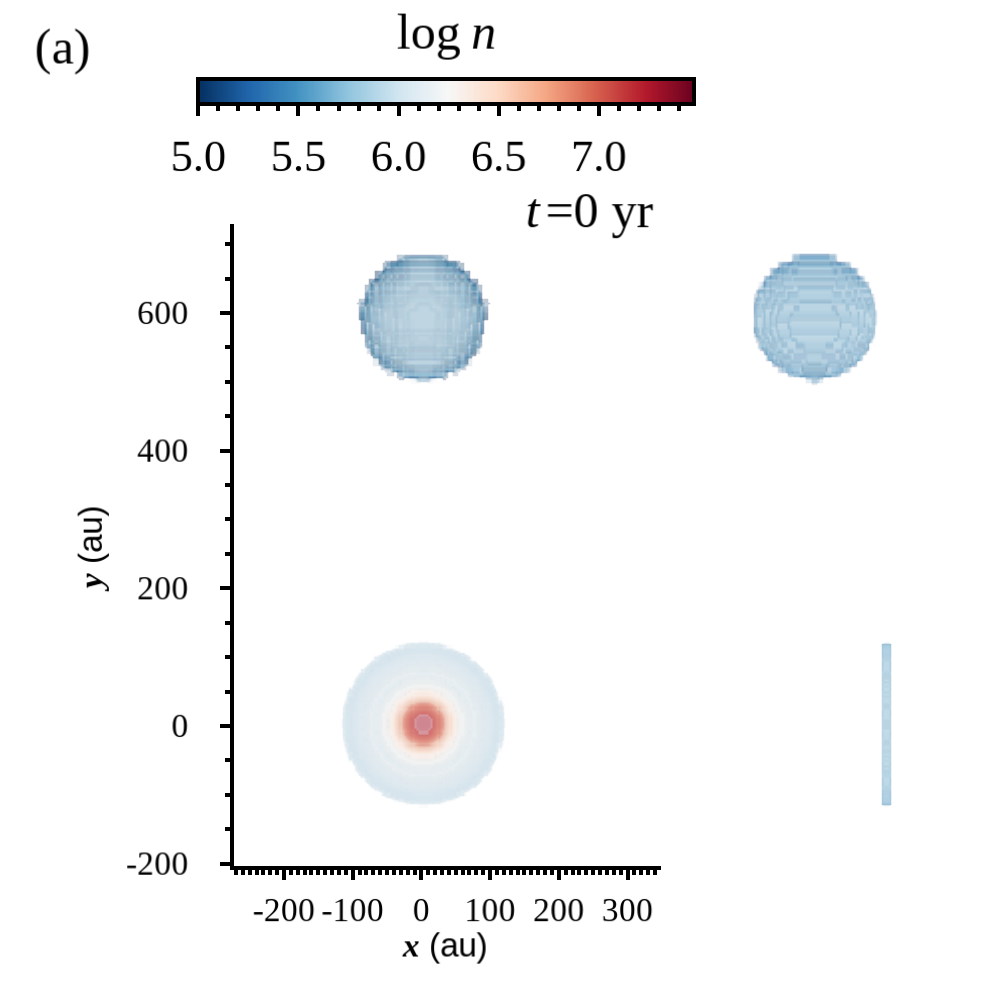}
\includegraphics[width=\columnwidth]{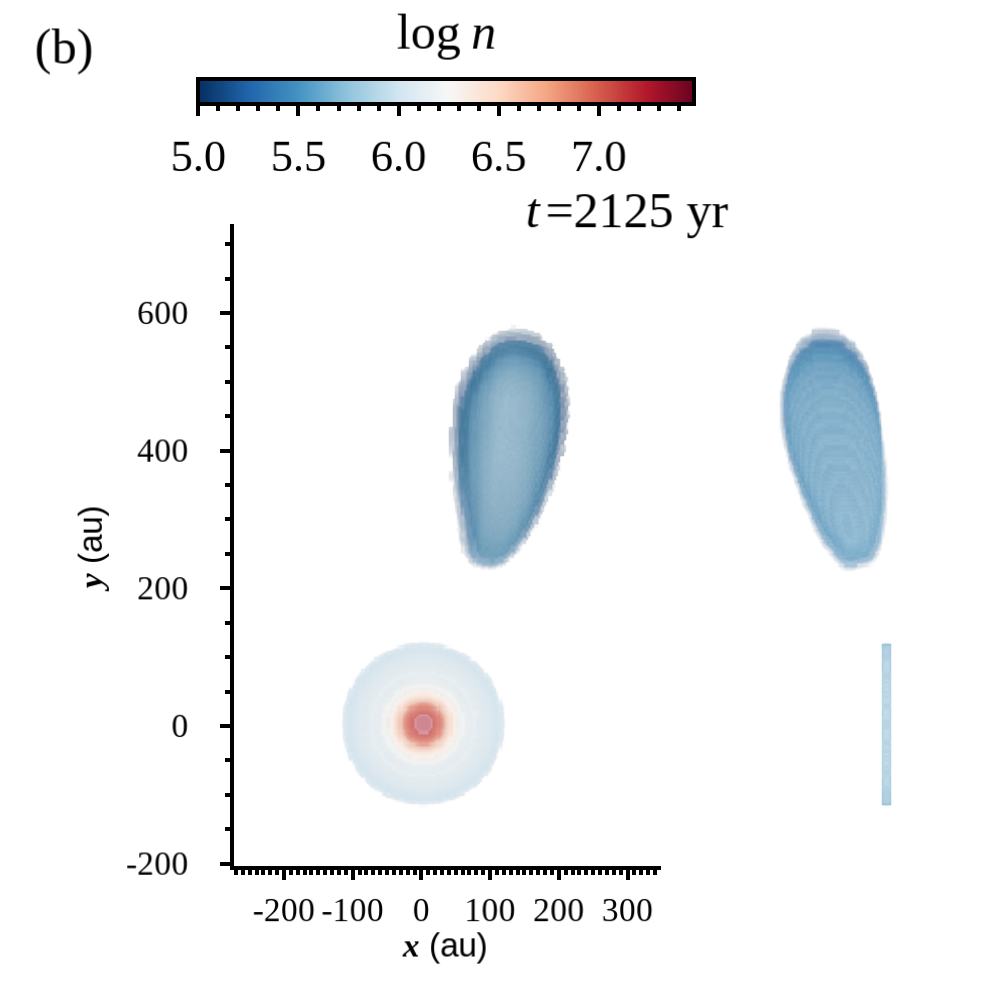} \\
\includegraphics[width=\columnwidth]{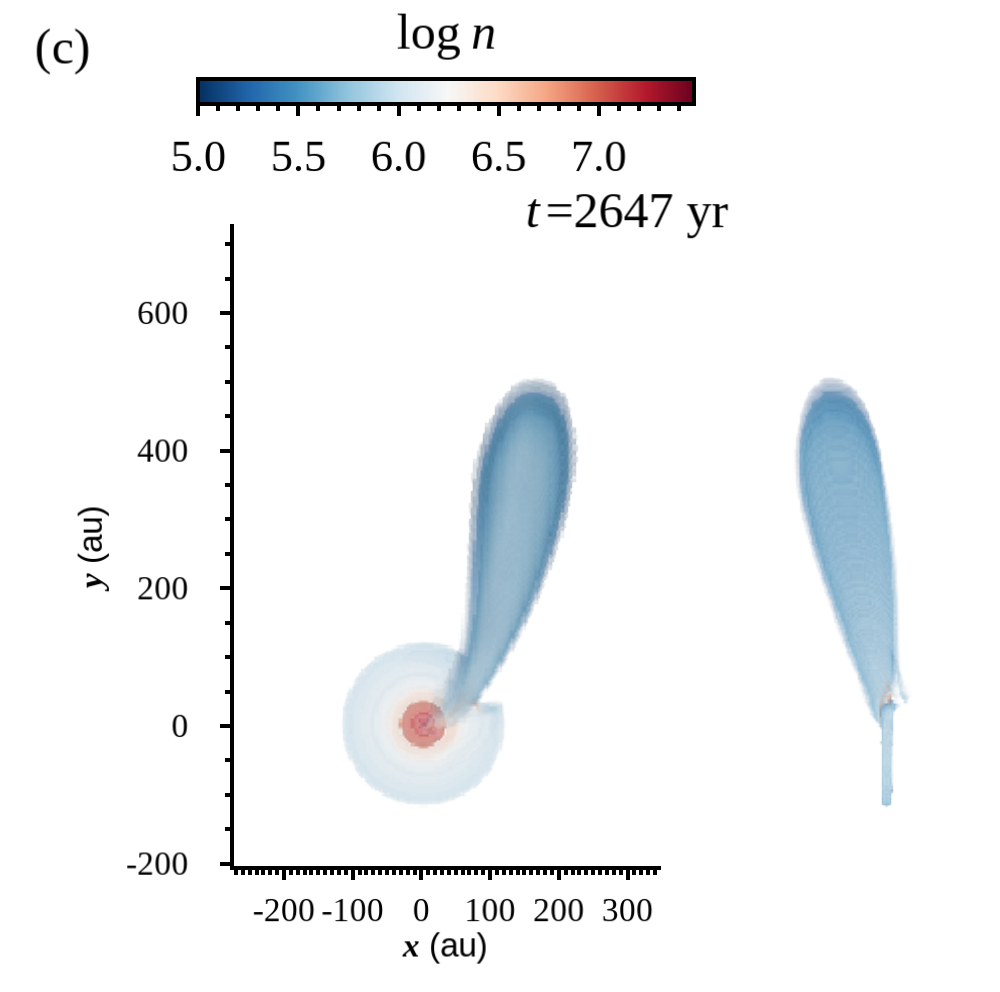}
\includegraphics[width=\columnwidth]{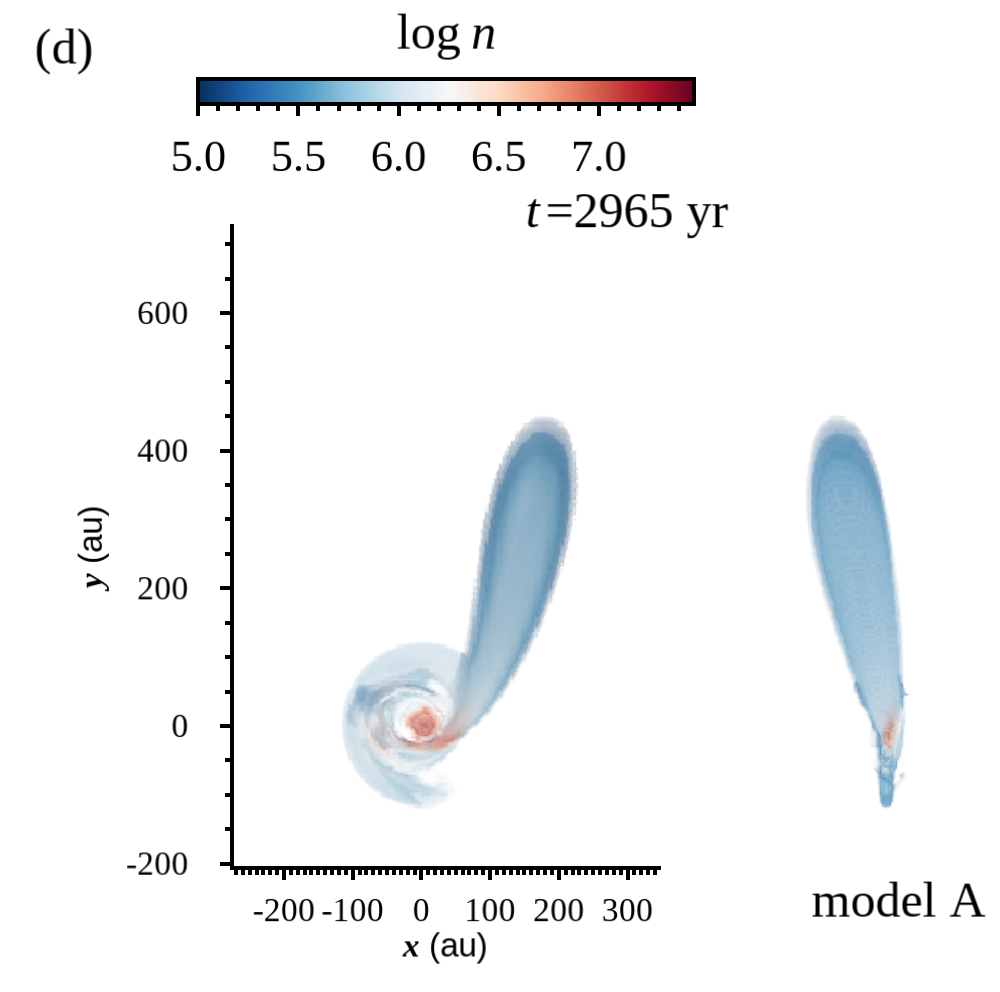}
\caption{Each panel denotes the cloudlet and disk in model A at a given epoch with top (left) and side (right) views. 
The color denotes the number density of molecules. The movie associated with this figure
includes stages up tp $ t \le 4859 $ yr and shows the disk evolution after impact}.\label{fig:A-rho}
\end{figure*}

Figure \ref{fig:A-rho} shows the infall of cloudlet in model A in
a series of snapshots. Each panel denotes the cloudlet and disk by
volume rendering. The color denotes the number density of the molecules, $ \rho |c| / \bar{m} _{\rm c} $, for the region of $ | c | \ge 0.5 $. The viewing angle is specified by
$ \varphi _{\rm obs} = 5.6 ^\circ $ and $ \chi = 45 ^\circ $
so that each panel mimics the observational image of DG Tau on the sky plane.
The warm gas ($ |c| < 0.5 $) is less dense than the colorbar threshold and invisible in the panels, 
though it fills the vast space outside of the disk and cloudlet.

At $ t = 2125~\mbox{yr} $, the cloudlet is prolate and
its head is $\sim$300 au away from the star.  
The radial motion dominates over rotation at this stage.
The acceleration in the radial direction and elongation are due to gravity and tidal force, respectively. As mentioned
in \S 2, we fixed the disk and its motion at the initial values.
We followed the change in the region of $ r \le R _{\rm d} ^\prime $
and $ |z| \le  Z _{\rm d} $ from $ t = 2337~\mbox{yr} $. At $ t = 2473~\mbox{yr}$,
the head of the cloudlet is close to the disk outer edge. The maximum density of the cloudlet is $ 1.25 \times 10 ^6~\mbox{cm} ^{-3}$ and only slightly higher than the initial value.  The pressure of the surrounding warm gas is nearly uniform in the region far from DG Tau ($ \sqrt{r ^2 + z ^2} > 120~\mbox{au}$).
Thus, the cloudlet does not shrink or expand appreciably, though the shape is highly elongated like streamer.
Finally, at $ t = 2499~\mbox{yr} $, the head of the cloudlet hits the disk from the upper surface. The main part of the cloudlet follows the head to form an accretion streamer.  

\begin{figure}
\centering
\includegraphics[width=0.35\textwidth]{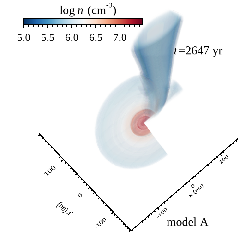} 
\caption{A zoom-in view of the disk and streamer in Model A at $ t = $ 2647~yr. The SW quarter of the disk is removed to show the vertical structure.
\label{fig:A-disk}}
\end{figure}

\begin{figure}
\centering
\includegraphics[width=0.48\textwidth]{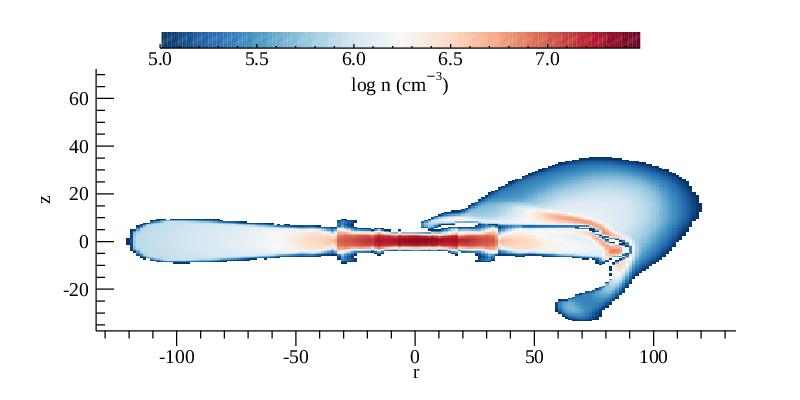} 
\caption{The cross section view of the disk and streamer shown in Fig.~\ref{fig:A-disk}.
\label{fig:A-disk-v}}
\end{figure}

Figure~\ref{fig:A-disk} shows a zoom of the stage shown in the lower right panel of Figure~\ref{fig:A-rho}, when the streamer impacts the disk. Figure \ref{fig:A-disk-v} shows the same stage as a cross section at $ x = 0 $. The northern side of the disk is deformed by the accretion.   
We see the shock-compressed gas at the intersection of the cloudlet and disk. We expect that 
SO molecules are formed in the shock due to dust grains sputtering and the release of S into the gas phase.
We also see the spiral shock waves in the disk. Since the disk is rotating around DG Tau, any dynamical impact tends to result in trailing spirals.
In model A, the streamer does not penetrate the disk but instead deflects at the surface. We note that the cloudlet and disk have almost the same pressure in our modeling. Both of them are in pressure equilibrium with the warm gas.
The disk is set to have twice higher temperature, and accordingly a lower density, than the cloudlet in model A. However, we do not claim that the disk gas is less dense than the cloudlet.  If we had taken account of the vertical temprature gradient, the density should be higher around the midplane.
We think that our model overestimates the impact of the streamer on the disk.

\begin{figure}
    \includegraphics[width=\columnwidth]{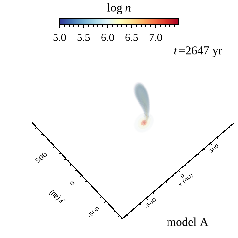}
    \caption{The bird's eye view for the stage at $ t~=~2647~\mbox{yr}$ in model A. The color denotes the number density of the molecules. The associated animation shows time evolution of model A by the bird's eye view from $ t = 0~\mbox{yr}$ to 4762~yr.
    \label{fig:modelA_birdeye}}
\end{figure}

Figure~\ref{fig:modelA_birdeye} and associated animation show the infall of cloudlet by the bird's view from $ t $ = 0~yr to 3099~yr. The viewing angle is set so that the disk rotation axis is 35$^\circ$ inclined with respect to the line of sight and the $ x $-axis lies on the position angle of PA~=~315$^\circ$. The cloudlet evolves into a structure elongated toward the North of DG Tau.  We use the same viewing angle for the Mock observation.

The accretion streamer should last for a few thousand years. We can estimate the duration from the initial size of the cloudlet. The tail is 180 au behind from the head and the initial velocity is $ 0.21~\mbox{km~s}^{-1} $.  This means that the delay of the tail is $ 2.7 \times 10 ^3~\mbox{yr} $.  

The impact of the streamer disrupts a part of the disk and leaves a significant imprint. The disk shows asymmetric features such as arms and spirals at least for several thousands years (see the animation associated with Figure \ref{fig:A-rho}). The disturbance may evolve into multiple rings after several disk rotations as demonstrated by \cite{demidova22}.  However, our model does not include any physical dissipation processes, and long-time evolution is beyond the scope of this paper.   

\subsection{Model B}

We have constructed model B to examine the effects of the initial distance. The cloudlet is 1200 au away from DG Tau at the initial stage of model B.  We have reduced the spatial resolution to cover a larger computational box and reduce computational cost. Since the initial distance is twice as large as the other models, the timescale for the cloudlet capture is three times longer. 

\begin{figure*}
\includegraphics[width=\columnwidth]{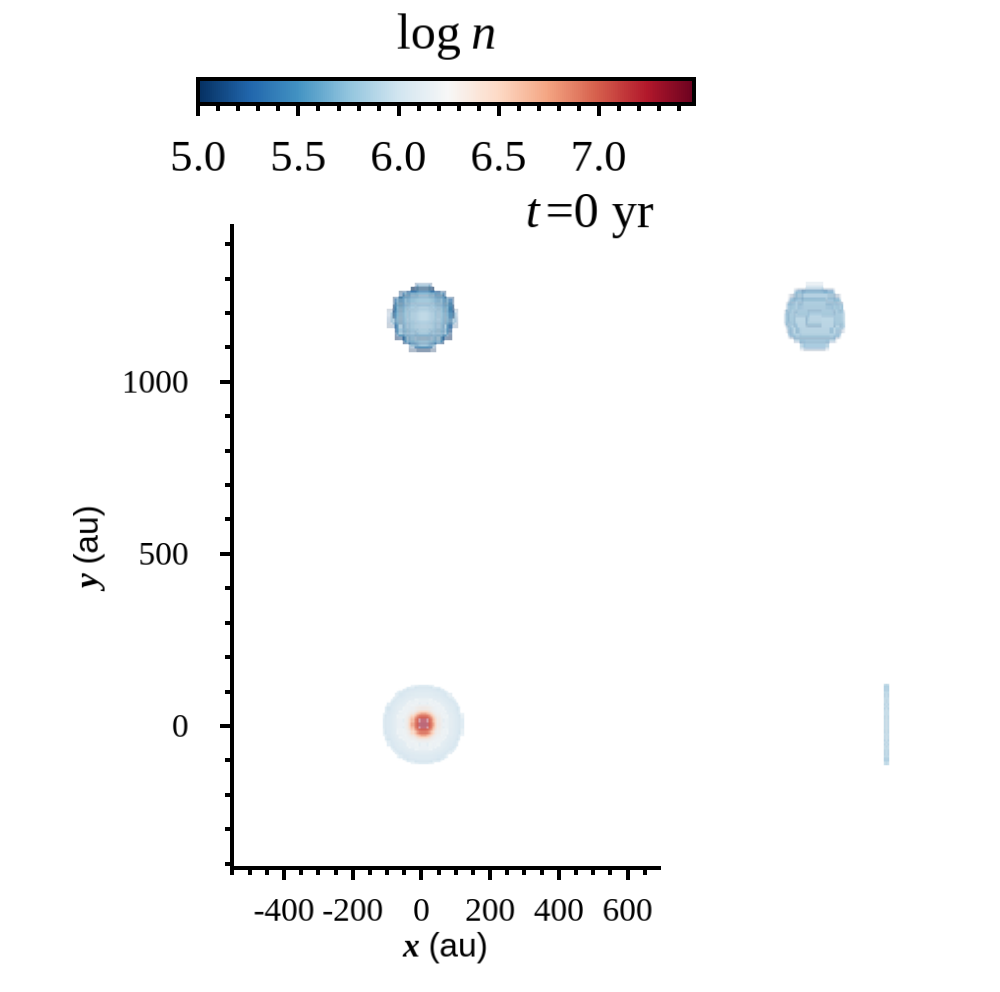}
\includegraphics[width=\columnwidth]{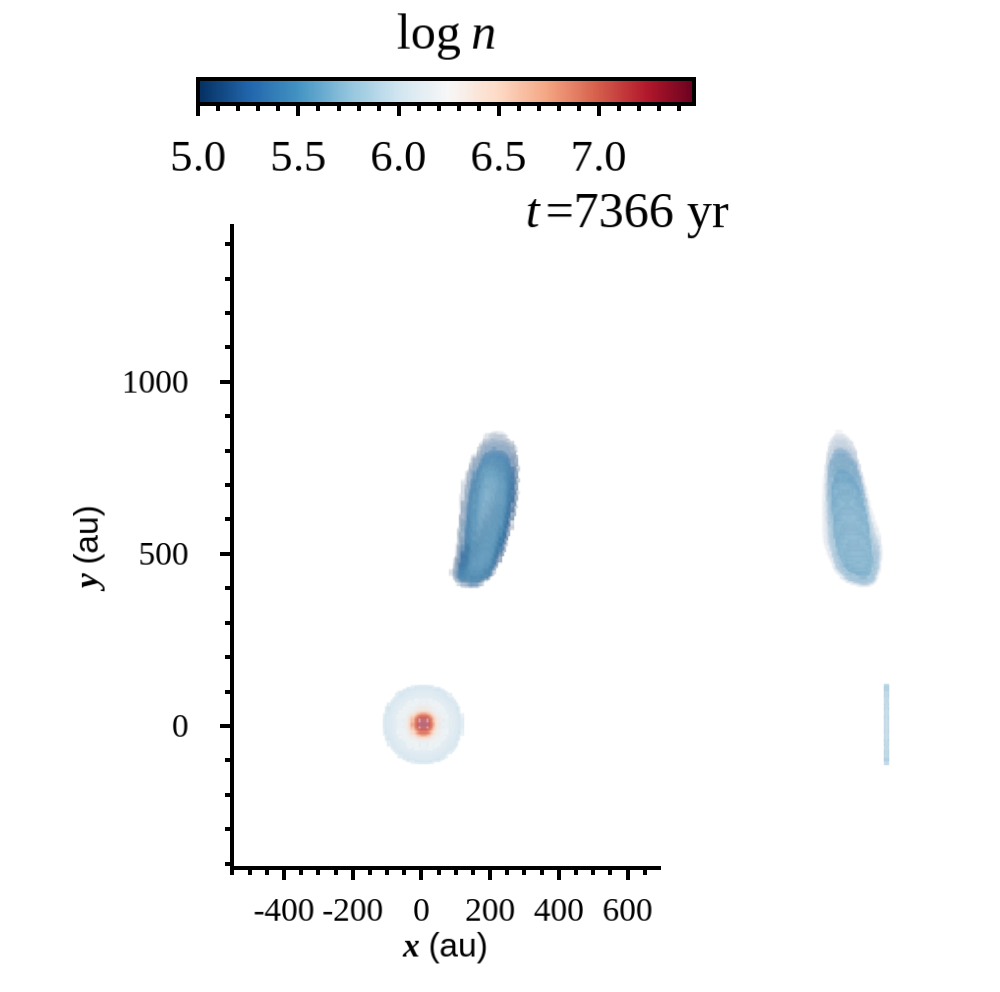} \\
\includegraphics[width=\columnwidth]{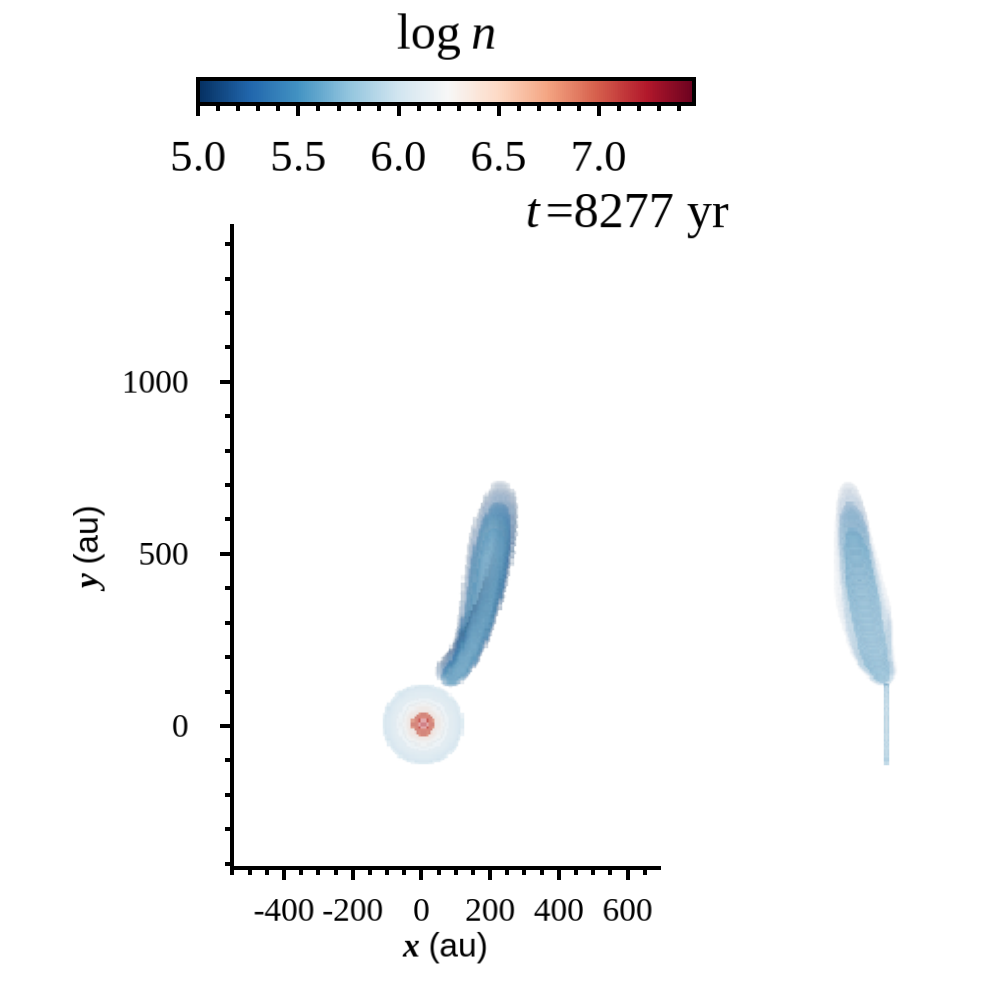}
\includegraphics[width=\columnwidth]{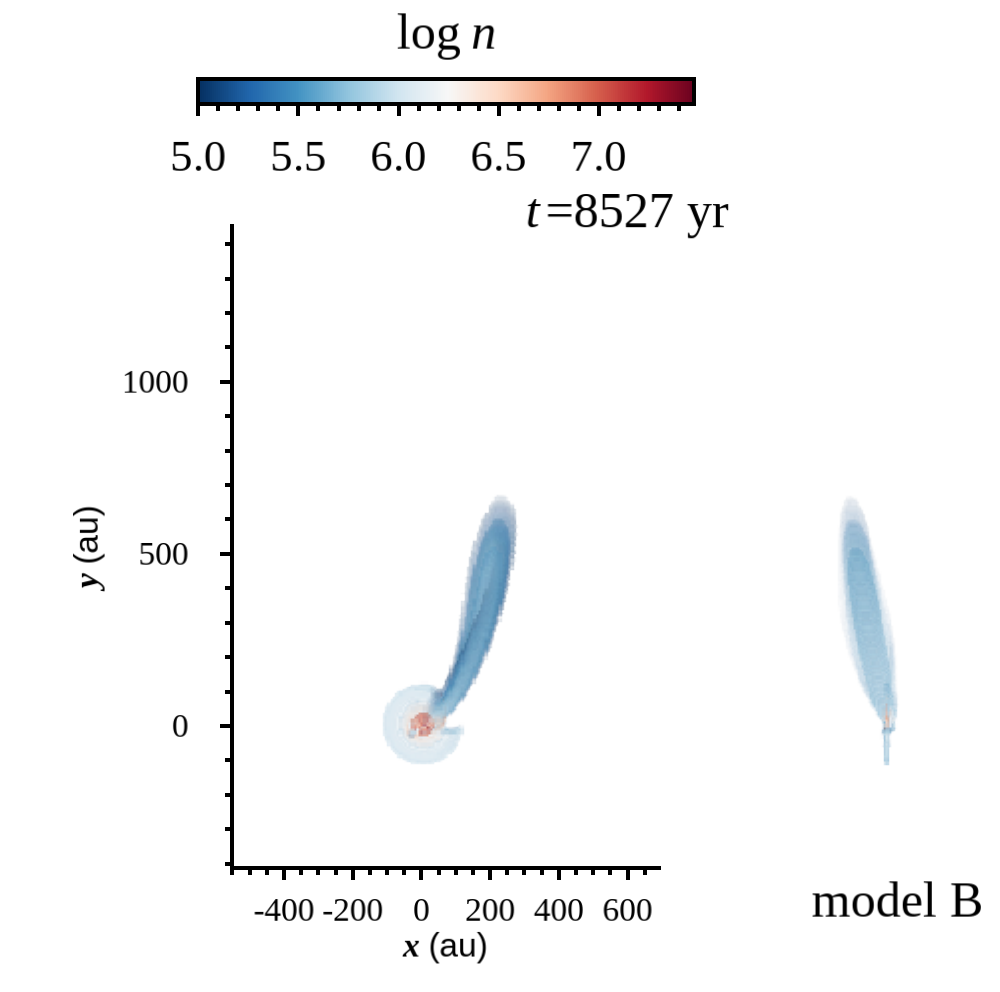}
\caption{The same as Fig.~\ref{fig:A-rho} but for model B. See the animation associated with this figure for further details of the change in the cloudlet form.\label{fig:B-rho}}
\end{figure*}

Figure~\ref{fig:B-rho} shows the capture of the cloudlet in a series of snapshots. While the cloudlet is spherical at the initial stage ($ t = 0$~yr), it changes first to prolate and later forms into a streamer. The cloudlet is elongated toward DG Tau at $ t = 7366 $~yr, though it is still far from DG Tau.

We stopped following the evolution at $ t = 8661~\mbox{yr}$ in model B. We could not follow the further evolution because of the lower spatial resolution. At this stage, the head of the streamer is immersed within the disk but the remaining part extends outside the disk.  

The cloudlet is elongated more in model B.  The elongation is due to the initial lower velocity. If the initial velocity further lower, a smaller cloudlet can evolve into a streamer. 

\subsection{Model C}

We have constructed model C to examine the case in which the clouldlet follows a parabolic orbit. The cloudlet is spherical and 600 au away from DG Tau at the initial stage, $ t = 0 $~yr.
The initial velocity of the cloudlet is $ \sim 1.4~\mbox{km~s} ^{-1} $ though it is not uniform in the cloudlet. This infall velocity is higher than the sound speed of the cloudlet while it is lower than that of the warm gas.

\begin{figure*}
\includegraphics[width=\columnwidth]{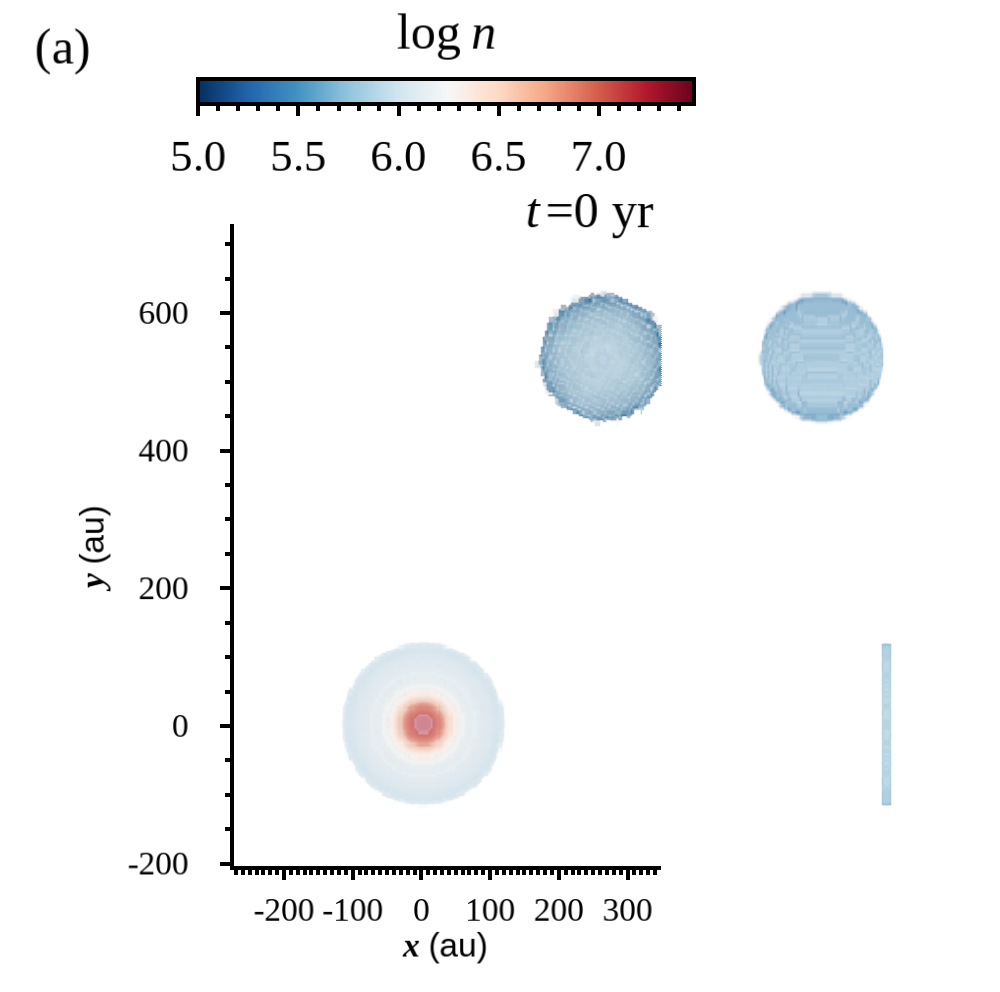}
\includegraphics[width=\columnwidth]{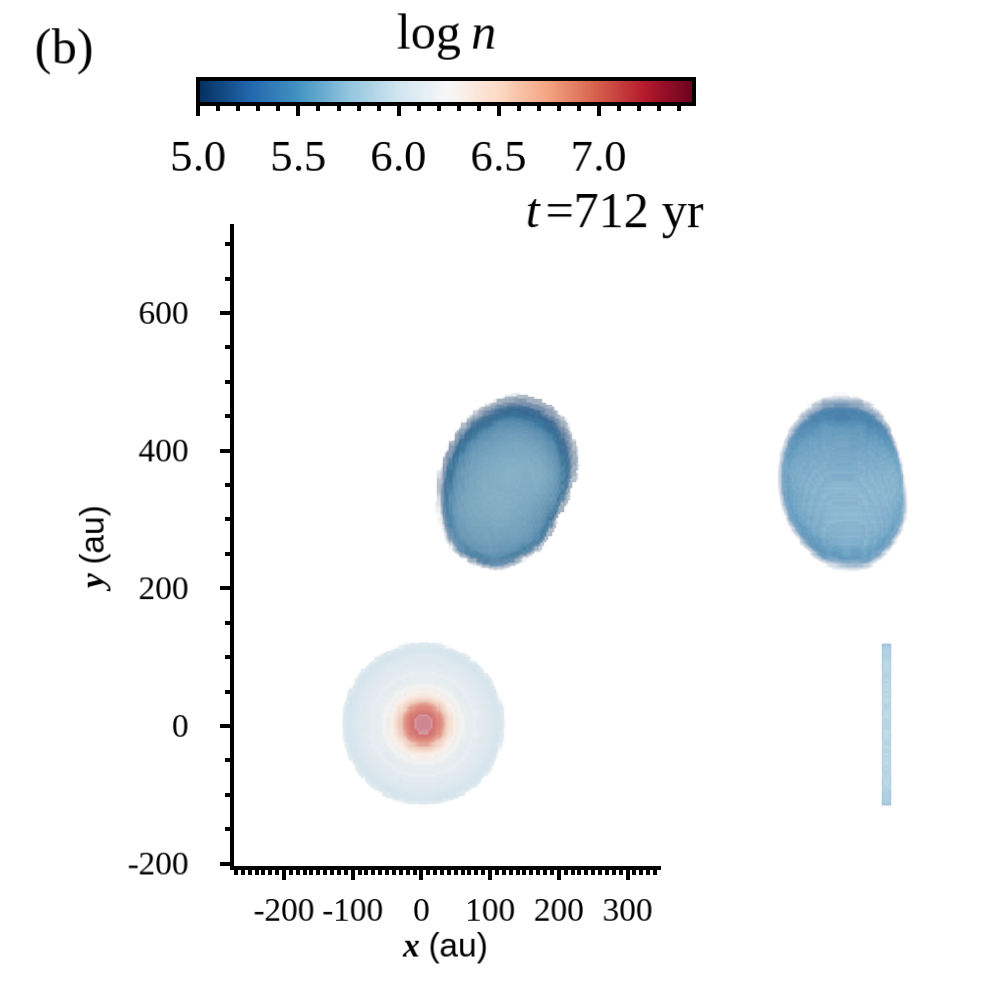} \\
\includegraphics[width=\columnwidth]{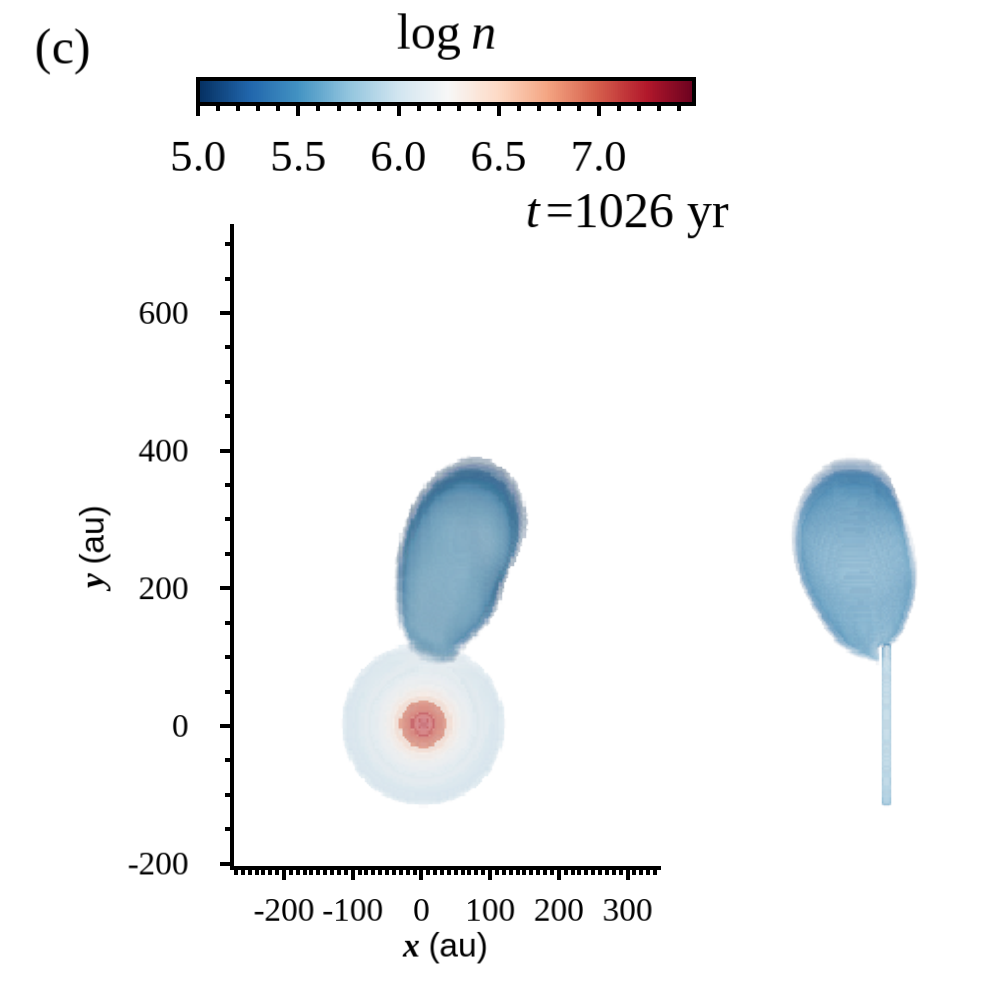}
\includegraphics[width=\columnwidth]{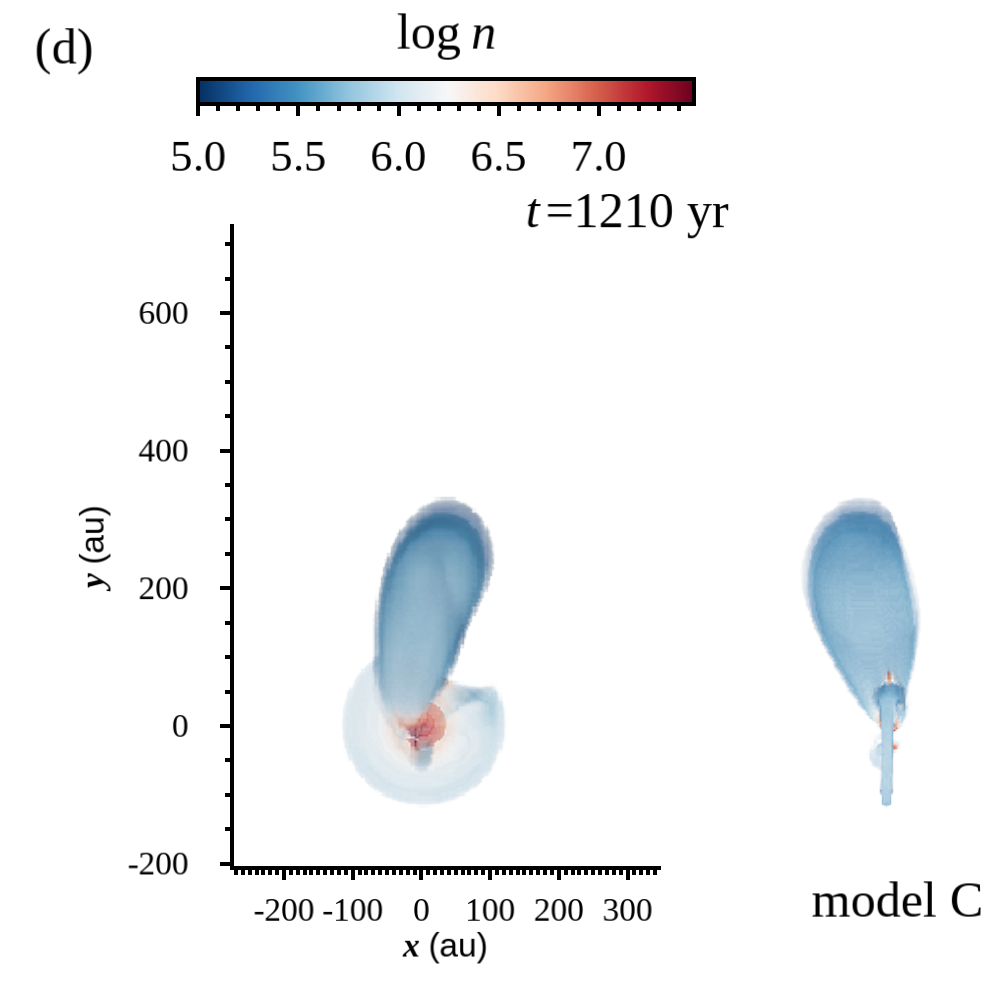}
\caption{The same as Fig.~\ref{fig:A-rho} but for model C. See the animation associated with this figure for further details of the change in the cloudlet form. \label{fig:C-rho}}
\end{figure*}

Figure \ref{fig:C-rho} shows the infall of the cloudlet in a series of snapshots. The cloudlet is slightly streched in the radial direction but the elongation is less prominent. Model C fails to reproduce the streamer, i.e., an elongated structure. The orbit does not follow that of a bullet. The changes in the shape and orbit are due to the drag force working on the cloudlet. The drag force is roughly proportional to the square of the relative velocity between the cloudlet and warm gas. Another factor is the larger initial velocity. Since the drag force works only on the front side, the rear side catches up to the head. We examine the drag force in \S 5. The higher initial velocity also suppresses the elongation.

\section{Comparison with observations}

\subsection{CS Channel maps}

\begin{figure*}
    \centering
    \includegraphics[width=0.98\textwidth]{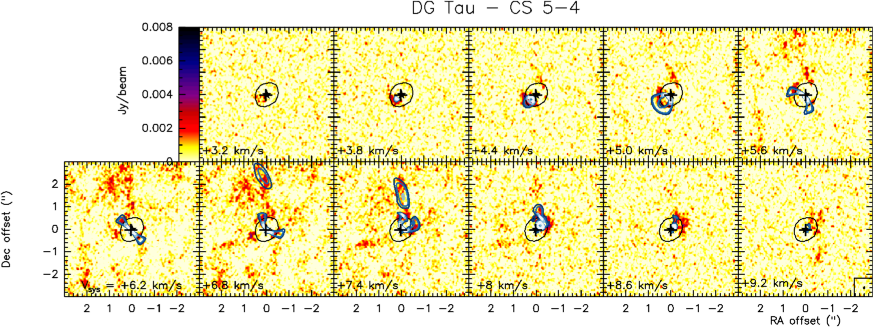}\\
    \includegraphics[width=0.98\textwidth]{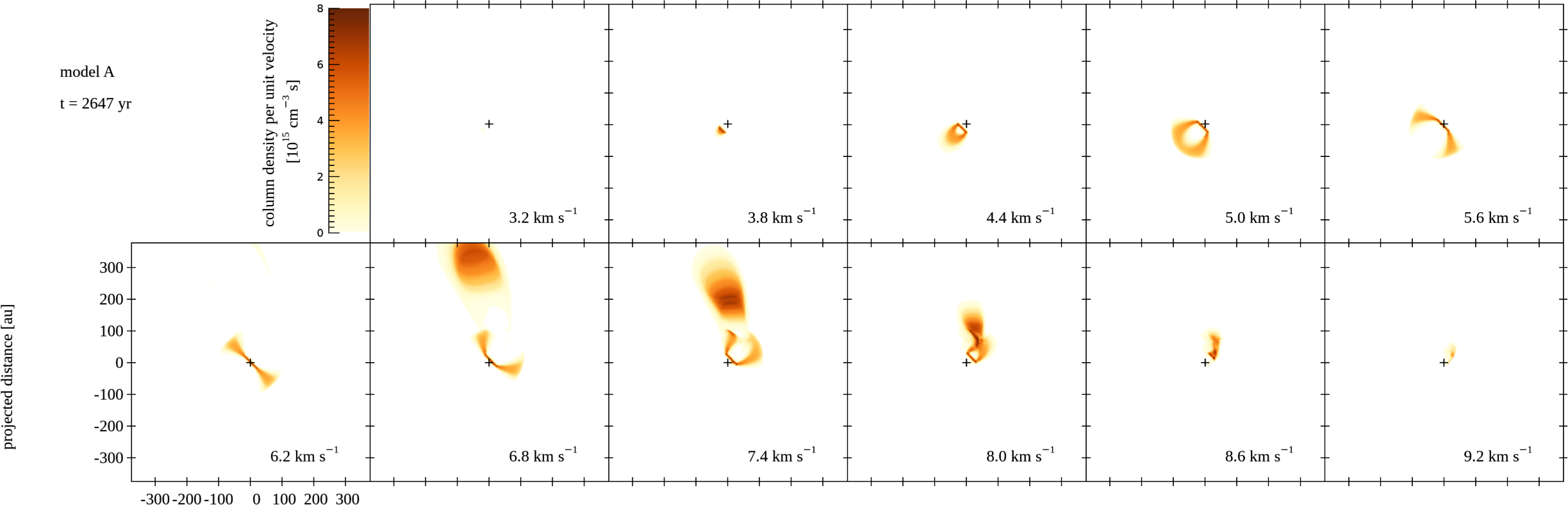}
 \caption{Comparison of CS ($5-4$) emission \citep{podio20} and mock channel maps for model A at $ t $ = 2647 yr.  The upper half denotes the line intensity by color and the column density per unit velocity by blue contours.  The lower half shows the latter by color.}  The color bars are given in the last row.
    \label{fig:channel}
\end{figure*}

The upper half of Figure~\ref{fig:channel} shows the channel maps of CS ($J=5-4$) line emission obtained with ALMA at a spatial resolution of $0.16^{\prime\prime} \times 0.13^{\prime\prime}$ and a spectral resolution of 0.6 km~s$^{-1}$ in the context of the ALMA-DOT program. Each panel covers an area of $6.0^{\prime\prime} \times 6.0 ^{\prime\prime} $ centered at the continuum peak. The details of the observations are given in \cite{garufi21, garufi22}. 
The CS line emission probes both the rotating gas in the disk and an accretion streamer north of the disk, which impacts on the NE disk side. Due to dust opacity effects in the inner disk region and/or due to excitation conditions, CS only probes the outer disk from a radius of $\sim 30$ au out to $\sim 130$ au. The streamer has two components, red-shifted and blue-shifted.

\subsection{Other Molecules and Dust Continuum}

\cite{podio19} reported a ring in the H$_2$CO emission line observed with ALMA. From the same data set, \cite{podio20} detected 1.3 mm continuum emission originating from the disk. The radius of the dust disk is $\sim 70$ au, and we find weak asymmetry in H$_2$CO or dust emission with the limited angular resolution data. The level of asymmetry is much lower than the disturbance shown at a later stage in our hydrodynamical models. 
\cite{ohashi23} have analyzed the archival data (2015.1.01268.S) to show that the dust continuum emission at 1.3 mm is smooth and does not show any substructures at a spatial resolution of 5 au. 
These observations therefore suggest that the disk midplane is for the most part undisturbed. In turn, this may indicate that either the entire disk has not been perturbed yet by the streamer, or that the disk is stiff and the dust at the midplane is not significantly affected by the streamer. We consider the former case when comparing our hydrodynamic model with the observations. The former possibility is supported by the localized nature of the SO emission. Unless the SO molecules are deposited back onto the dust grains within a few hundreds of years, the SO emission should form an arc extended along the gas stream, like in the case of HL Tau described by \citet{garufi22}.\\

The ALMA maps obtained by \cite{bacciotti18,podio19,podio20}
show that both the molecular emission (H$_2$CO and CS ($5-4$)) and the continuum emission at 0.8 and 1.3mm are enhanced in the NW side, i.e. in the region where the streamer impacts the disk
See Figures 1 and 3 from \cite{podio20}, Fig. B.1 from \cite{podio19}, and Fig. 2 from \cite{bacciotti18}. 
The lack of SO emission along the gas stream in the disk, as observed in the case of HL Tau \citep{garufi22}, could be due to the lower angular resolution and signal-to-noise ratio of the SO observations for DG Tau (beam size $\sim0.35\arcsec$ and rms $\sim0.7$ mJy beam$^{-1}$).

\subsection{Model A}

We compare model A with the ALMA observation since it gives the bet among our models. 
The third and fourth rows of Figure~\ref{fig:channel} are the mock channel maps based on the stage at $ t = 2647 $~yr in model A. Each map covers a square of 250~au~$\times$~250~au centered at DG Tau, which is
marked by the cross on each map. The spectral resolution is
set to be 0.6~km~s$^{-1}$ for comparison with the observations.

The channel maps show the rotating disk and the Northern streamer. We find a strong emission spot at 7.6~km~s$^{-1}$. This spot corresponds to the shock compressed cloudlet head.  Thus, it corresponds to the SO emission spot. Model A cannot reproduce the blue-shifted streamer component seen in the observations. See also Figure \ref{fig:modelA_birdeye} for the morphology of the cloudlet. It shows the elongated cloudlet colliding with the disk at the stage shown in the mock observation. The associated animation shows the formation of the streamer and evolution of the disk after the impact by bird's eye view.

\section{DISCUSSIONS}

A cloudlet approaching DG Tau transforms into a streamer 
under some circumstances.  As long as the drag force is weak,
the tidal force elongates the cloudlet, and the warm surrounding
gas compresses it from the side.  We examine the strength of the drag force.

Newton's law gives the drag force (resistance) acting on a body moving at velocity $ v $,
\begin{eqnarray}
F _{\rm drag} & = & C _{\rm D} S \rho _{\rm w}  \frac{v ^2}{2} , \label{drag}
\end{eqnarray}
where $ \rho _{\rm w} $, $ S $ denote the density of the ambient (warm) gas and the cross section of the body, respectively.  The symbol, $ C _{\rm D} $, denotes a non-dimensional number, and is evaluated to be $ \approx 0.4 $. Equation (\ref{drag}) is the same as Equation (7) of \cite{weidenschilling77}, who evaluated the drag force acting on a solid particle in the solar nebula.  The adopted value of $ C _{\rm D} $ is appropriate when the Reynolds number is very large, i.e., the body size is much larger than the mean free path of the gas molecule. The drag force not only decelerates but shortens the cloudlet since it works only on the front side.

We compare the drag force with the tidal force, since the latter works to elongate the cloudlet. It pulls the front side toward
the star and pushes the rear side away. When the cloudlet is a gas sphere of radius $ a _c $, the tidal force is estimated to be
\begin{eqnarray}
F _{\rm tidal} & = & \frac{\pi G M \rho _{\rm c} a _c {}^4}{r _c {}^3} . 
\end{eqnarray}
Here, the tidal force is defined as the difference between the force extending the head forward and that extending the tail backward.
The drag force works in the direction against the cloudlet motion while the tidal force works in the direction towards the star.
Though the drag force has an azimuthal component, the radial part is dominant except for near the periastron and apoastron since we consider parabolic or highly eccentric orbits.  Thus, we assume it has only the radial component for simplicity.
Then the ratio of these forces is expressed as
\begin{eqnarray}
\frac{F _{\rm tidal}}{F _{\rm drag}} & = &
\frac{1}{C _{\rm D}} \frac{\rho _{\rm c}}{\rho _{\rm w}}
\left( \frac{a _c}{r _c} \right) ^2
\frac{2 G M}{r _c v ^2} . \label{force_ratio}
\end{eqnarray}
The tidal force is dominant when (1) the cloudlet is dense 
($ \rho _{\rm c} \gg \rho _{\rm w}) $, (2) the velocity is low, and (3) the viewing angle of cloudlet from the star is relatively large. 
Note that the velocity is normalized by the gravity in Equation (\ref{force_ratio}). 

In model C, the drag force is dominant over the tidal force mainly because the initial velocity is high.  In the other models, the tidal force is dominant since the initial velocity is low ($ v ^2 \ll GM/r _c $). The drag force compresses the head of the cloudlet and the compression propagates in the cloudlet at the sound speed. When the cloudlet's motion is subsonic, the compression propagates through the cloudlet and the shape changes little.  Otherwise, the cloudlet has a highly compressed layer near the head as shown in model C. 

Once the cloudlet is elongated, the elongation continues. The elongation reduces the drag force through the reduction in the cross section, and enhances the tidal force. As shown in model B, 
an elongated cloudlet transforms into a streamer. 
Equation (\ref{force_ratio}) provides a constraint on the early phase of streamer formation.

As far as we know, direct observations of the ambient warm gas have been not yet reported, although it is expected to be present \citep[see, e.g., Figure 1 of][]{dutrey14}. The morphology of streamers could give indirect evidence of the warm gas and a constraint on the temperature.

The impact of the cloudlet depends on the inclination of the
orbital plane and disk thickness.  When the orbit is coplanar with
the disk, the cloudlet collides with the outer edge of the disk.
When the orbit is inclined, the cloudlet hits the disk surface.
In our model, $ \theta _0 $ specifies the orbital plane.  
When $ (H/r) \tan \theta > 1 $, the orbit is coplanar, where
$ H $ and $ r $ denote the disk height and radius, respectively.
Since the SO emission appears at a radius of 50 au in DG Tau, the streamer's orbit should be inclined with the disk.  We assumed the inclination of 10$^\circ$ in our models in accordance with \cite{garufi22}. The ratio of the height to radius of the CS emitting layer is estimated to be $ H/r \sim 0.09 $ by \cite{garufi21} from the asymmetry between the near and far sides of the disk.
Model A is consistent with these observations.  
The orbital plane of the cloudlet should be located above the disk surface in the northeast.

\begin{figure}
\begin{center}
\includegraphics[width=0.47\textwidth]{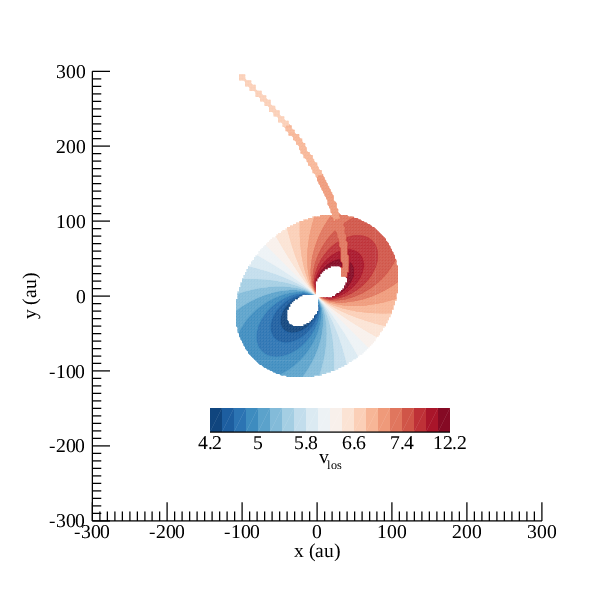}    
\end{center}
    \caption{Synthetic moment 1 map for the streamer and disk of DG Tau.  
    The streamer is assumed to be a part of the elliptic orbit.
    The orbital element is the same as that for the cloudlet center
    of model A. 
    The streamer is red-shifted.
    \label{fig:bullet-v_los}}
\end{figure}

Our model implies the possibility that we could detect the velocity gradient along the streamer if the spectral resolution of the observations had been higher. The velocity gradient is also shown in the ballistic modeling of \cite{garufi22}. Applying the ballistic approximation, we have made a moment 1 (intensity-weighted mean velocity) map, Figure~\ref{fig:bullet-v_los}. The colored curve denotes the cloudlet's orbit projected on the sky plane while the colored ellipse denotes the rotating disk. The orbital elements are the same as those of model of A1.
The streamer is accelerated in the region where the projected distance is less than $ r _{\rm proj} \la 100~\mbox{au}$. The projected orbit is curved only near the periastron. The accretion streamer indicates that the specific angular momentum is already low in the region of $ r _{\rm proj} > 100~\mbox{au}$.

The SO emission is a tracer of recently shocked, compressed gas as these molecules are rapidly formed in shocks due to the release of sulphur from the dust grain mantles \citep[e.g.,][]{neufeld89a,neufeld89b}.
\cite{vangelder21} estimated the lifetime of the SO molecule to be
deposited back onto the dust as several hundreds of years in their modeling accretion shocks at the disk-envelope interface.  We expect that the SO molecules flow out from the impact point since the gas disk is rotating.
The Keplerian rotation velocity is 3.5~km~s$^{-1}$ at the distance of 50 au for the assumed mass of 0.7~M$_\odot$.
This velocity is equivalent to 74 au per century. The localization of the SO emission on the disk suggests two possibilities: (1) the impact of the streamer is a recent event, (2) the lifetime of SO molecule is shorter in DG Tau than expected in models.
This would be consistent with the observations that the disk of DG Tau has relatively minor asymmetries in the dust emission \citep[e.g.,][] {podio19}. Alternatively, as noted above, the lack of SO spiraling through the disk may be due to the lower resolution of the observations or fainter emission in DG Tau compared to HL Tau.

We discuss the dependence of our model on the disk gas temperature since we neglected the temperature gradient in the disk.  The temperature should be lower near the mid plane and in the outer part of the disk.  
When the temperature is lower, the gas density is higher for a given surface pressure. If the disk is more massive, the larger inertia should soften the dynamical impact.  

As summarized by \cite{pineda23}, many young stellar objects show asymmetric accretion features feeding disks, though the morphologies have variety in shape and dynamics. The form may reflect the environment and evolutionary stage. HL Tau shows an accretion streamer similar to that of DG Tau. Both DG Tau and HL Tau classical T Tauri stars, and both have relatively symmetric disks.  More evolved stars are often associated with asymmetric extended disks \citep[e.g.,][]{boccaletti20}.  On the other hand, some Class 0 sources such as TMC 1A often show asymmetry but not streamers \citep{aso15,sakai16}. DG Tau is also associated with another type of asymmetry, the southern arc \citep{garufi22}. Because the southern arc has a large velocity offset at a large projected distance, it is unbound to DG Tau and unlikely to be an accreting streamer.

\section{SUMMARY}

We have shown that our hydrodynamic model can reproduce an accretion streamer infalling to DG Tau by assuming the following conditions.  

First, we considered a warm neutral medium of atomic gas to prevent expansion of the streamer.  We can see the warm neutral medium neither in the molecular emission lines nor in the dust continuum. Though the density is low, it has a gas pressure comparable to that of the molecular gas because of the relatively high temperature.  

Second, the infalling cloudlet has small radial and rotation velocities at the initial stage.  If the rotation velocity is large, the cloudlet cannot reach the disk because of the large centrifugal force.  If the radial velocity is large, the drag force prevents the elongation of the cloudlet. 

Third, the inclination of the cloudlet's orbit should be higher than the opening angle of the disk, $ z / r $. Otherwise, the spot of SO emission at the impact zone should appear at the disk outer edge.

Fourth, the accretion streamer should have reached the disk only recently ($ < 10 ^3~\mbox{yr}$) in DG Tau.  Otherwise, we would see more dramatic perturbations in the observations of the disk.

\section*{Acknowledgements}

We thank an anonymous reviewer for the valuable comments.
This work was supported by JSPS KAKENHI grant Nos. JP18H05222, JP19K03906, JP20H05847, and JP20H00182. CC and LP acknowledge 
the PRIN-MUR 2020 MUR BEYOND-2p (Astrochemistry beyond the second period elements, Prot. 2020AFB3FX).
D. S.-C. is supported by an NSF Astronomy and Astrophysics Postdoctoral Fellowship under award AST-2102405.

\section*{Data Availability}

This paper makes use of the following ALMA data: ADS/JAO.ALMA\#2015.1.01268.S. ALMA is a partnership of ESO (representing its member states), NSF (USA) and NINS (Japan), together with NRC (Canada), MOST and ASIAA (Taiwan), and KASI (Republic of Korea), in cooperation with the Republic of Chile. The Joint ALMA Observatory is operated by ESO, AUI/NRAO and NAOJ.




\bibliographystyle{mnras}
\bibliography{DGTau} 




\bsp	
\label{lastpage}
\end{document}